\documentclass[aps,prb,twocolumn,preprintnumbers,amsmath,amssymb,superscriptaddress,floatfix]{revtex4}
\usepackage{graphicx}

\begin{document}
\title{Synthetic electric fields and phonon damping in carbon nanotubes and graphene}

\author{Felix von Oppen}
\affiliation{Institut f\"ur Theoretische Physik, Freie Universit\"at
Berlin, Arnimallee 14, 14195 Berlin, Germany}

\author{Francisco Guinea}
\affiliation{Instituto de Ciencia de Materiales de Madrid, CSIC, Cantoblanco E28049 Madrid, Spain}

\author{Eros Mariani}
\affiliation{Institut f\"ur Theoretische Physik, Freie Universit\"at
Berlin, Arnimallee 14, 14195 Berlin, Germany}

\begin{abstract}
Within the Dirac theory of the electronic properties of graphene,
smoothly varying  lattice strain affects the Dirac carriers through
a synthetic gauge field. For static lattice strain, the gauge field
induces a synthetic magnetic field which is known to suppress weak
localization corrections by a dynamical breaking of time reversal
symmetry. When the lattice strain is time dependent, as in
connection with phononic excitations, the gauge field becomes time
dependent and the synthetic vector potential is also associated with
an electric field. In this paper, we show that this synthetic
electric field has observable consequences. We find that the Joule
heating associated with the currents driven by the synthetic
electric field dominates the intrinsic damping, caused by the
electron-phonon interaction, of many acoustic phonon modes of
graphene and metallic carbon nanotubes when including the effects of
disorder and Coulomb interactions. Several important consequences
follow from the observation that by time-reversal symmetry, the
synthetic electric field associated with the vector potential has
opposite signs for the two valleys. First, this implies that the
synthetic electric field drives charge-neutral valley currents and
is therefore unaffected by screening. This frequently makes the
effects of the synthetic vector potential more relevant than a
competing effect of the scalar deformation potential which has a
much larger bare coupling constant. Second, valley currents decay by
electron-electron scattering (valley Coulomb drag) which causes
interesting temperature dependence of the damping rates. While our
theory pertains first and foremost to metallic systems such as doped
graphene and metallic carbon nanotubes, the underlying mechanisms
should also be relevant for semiconducting carbon nanotubes when
they are doped.

\end{abstract}

\maketitle
\section{Introduction}
It is one of the remarkable aspects of the Dirac description of the
low-energy electronic properties of graphene \cite{Netal04,Netal05}
that both disorder and lattice strain give rise to synthetic gauge
fields.\cite{NGPNG09,GGV92,SA02b,Metal06,M07} If the synthetic
vector potential ${\bf A}({\bf r})$ is due to static  disorder or
static lattice distortions, its presence affects the electronic
dynamics through an effective magnetic field $e{\bf
B}=\nabla\times{\bf A}$ which, by time reversal symmetry, points in
opposite directions at the two Dirac points of the electronic
dispersion. Consequences of this effective magnetic field have been
widely studied in the
literature.\cite{Metal06,MG06,GKV08,MO08,GHL08} Additional physics
arises when the synthetic vector potential is caused by {\it
time-dependent} distortions such as phonons. In this case, ${\bf A}$
becomes time dependent and will generate not only a magnetic but
also an effective electric field $e{\bf E}=-\partial{\bf A}/\partial
t$.

In this paper, we show that these synthetic electric fields have
observable  consequences. Consider a low-energy phonon mode of
graphene or carbon nanotubes (CNT). The phonon is associated with a
synthetic electric field which, when the system is metallic, drives
currents. The dissipation (Joule heating) associated with these
currents causes damping of the phonon mode. We find that frequently,
the synthetic electric fields are directly responsible for the
damping of phonon modes in metallic carbon nanotubes and graphene.
In the clean limit, this damping mechanism is equivalent to
dissipation by electron-hole pair creation. In fact, we find that we
reproduce corresponding recent results for the radial breathing mode
of clean carbon nanotubes.\cite{Setal08b} Approaching the problem
from the point of view of the synthetic electric fields allows us to
calculate damping rates including the effects of disorder and
electron-electron interactions which we find to be significant.
Moreover, we  find appreciable damping rates even for those phonon
modes of carbon nanotubes for which damping by electron-hole pair
creation is not effective due to the large discrepancy between the
electron and the phonon velocities. As a result, we expect damping
by the electron-phonon interaction to dominate over other mechanisms
such as phonon-phonon coupling\cite{MEG09} over a wide range of
parameters.

Searching for observable consequences of the synthetic electric
field is  complicated by the fact that lattice distortions do not
only induce a vector potential ${\bf A}$ but also a scalar
potential\cite{SA02b} $\phi({\bf r},t)$ which leads to an additional
electric field $-\nabla\phi$. Within a tight-binding approach, the
synthetic vector potential is associated with changes in the hopping
amplitude caused by changes in the bond length. The scalar potential
is a deformation potential which arises from local dilation or
compression of the lattice. In fact, estimates\cite{SA02b} suggest
that the bare coupling constant of the scalar potential is about an
order of magnitude larger than that of the vector potential.

Nevertheless, we find that the damping of most (but not all)
low-energy phonon modes of graphene and carbon nanotubes is
dominated by the vector potential. This is a consequence of two
important qualitative differences between the electric fields
associated with the scalar and vector potentials: (i) While the
scalar field affects both valleys in the same manner, the sign of
the vector potential is opposite for the two valleys. As a result,
the currents driven by the electric fields are charge currents for
the scalar potential, but valley currents for the vector potential.
Therefore, the valley electric fields due to the vector potential
will not be subject to screening as they do not induce any charge
densities. Since in contrast, the electric fields arising from the
scalar deformation potential are screened, this may significantly
affect the relative importance of the two electric fields when the
system is metallic. (ii) The scalar potential is necessarily
associated with a longitudinal electric field whose direction is
parallel to the wavevector. In contrast, the vector  potential
generates a synthetic electric field which in general has both
longitudinal and transverse components. We find that over a wide
range of wavevectors, the transverse conductivity (and hence the
corresponding dissipation) is significantly larger than the
longitudinal conductivity.

The fact that the vector potential drives valley currents has
another  important consequence. While by momentum conservation,
electron-electron interactions cannot induce a decay of charge
currents, they do lead to a decay of valley currents by a process
which is analogous to spin Coulomb drag.\cite{Vig01,Fle01,Web05} We
find that this valley Coulomb drag mechanism leads to interesting
temperature dependence of the phonon damping which in many cases is
decreasing as temperature increases.

The damping mechanisms which we discuss apply most directly to
systems with  metallic behavior, i.e., metallic carbon nanotubes and
graphene. For this reason, we will not explicitly discuss
semiconducting CNT in this paper. It has recently been suggested
\cite{MEG09} that the intrinsic vibrational damping in
semiconducting CNT is dominated by nonlinear elastic effects.
However, it should be kept in mind that in practice, many
semiconducting CNT are doped and can hence exhibit a finite
conductivity. In this case, the damping mechanisms discussed in this
paper may well be relevant as well.

It is found experimentally that the vibrational relaxation time of
the  radial breathing mode is remarkably long, of the order of
several nanoseconds.\cite{LLKD04} This follows from
scanning-tunneling-microscope-based transport measurements, in which
the tunneling current excites the radial breathing mode, which
enables vibron-absorption processes at temperatures far below the
phonon frequency. We find that our results for the vibrational relaxation time
of the radial breathing mode are consistent with these
experiments.

Identifying parameters for which long vibrational relaxation times
can be  realized in CNT is also of much interest from the
perspective of nanoelectromechanical systems made of carbon
nanotubes\cite{LLKD04,RMPR06,Setal06,WPV06,Setal08b,HPWZ08,Letal09}
and graphene.\cite{Getal07b,Betal07,Getal08} For example, if vibrational
relaxation times are sufficiently long, these systems provide
access to a regime in which the transport current drives the
vibrational mode far out of thermal equilibrium. For strong
electron-vibron coupling, this nonequilibrium regime can be
characterized by self-similar avalanche transport.\cite{KO05,KOA06}
One of the most promising systems to observe this effect are suspended carbon
nanotube quantum dots for which strong electron-vibron coupling and
the associated Franck-Condon blockade have recently been confirmed
experimentally.\cite{Letal09}

This paper is organized as follows. In Sec.\ \ref{basic}, we
introduce the strain-induced scalar and vector potentials (Sec.\
\ref{strain}) and derive a general expression for the relation
between the phonon damping rate and the conductivity (Sec.\
\ref{PhononDamping}). We also illustrate the basic physical picture
in the context of the radial breathing modes of carbon nanotubes
(Sec.\ \ref{Picture}). Our approach allows us to include the effects
of disorder (Sec.\ \ref{disorder})  and of electron-electron
interactions (Sec.\ \ref{interaction}) on phonon damping. Finally we
conclude in Sec.\ \ref{conclusions}. Some calculational details are
relegated to an appendix.

\section{Basic formalism}
\label{basic}

\subsection{Strain-induced vector and scalar potentials}
\label{strain}

Elastic strains couple to electrons in graphene and carbon nanotubes
by inducing effective scalar and vector potentials into the
low-energy electronic Dirac Hamiltonian,\cite{NGPNG09}
\begin{equation}
{\cal H} =  \int d^2 {\bf r} \Psi^\dag ( {\bf r} )\left\{
v_F\left[ i \sigma_i ( \partial_i - A_i ) \right] + \phi \right\}\Psi ( {\bf r} ),
\end{equation}
where $\Psi ( {\bf r} )$ is the spinor which defines the electron,
$\sigma_i$ is a  Pauli matrix, $v_F$ is the Fermi velocity, and we
show the Hamiltonian defined for one of the two, $K$ and $K'$,
valleys in the Brillouin zone of graphene. Here we chose coordinates
such that the unit vectors of the graphene lattice can be written as
\begin{align}
{\bf a}_1 &= a \sqrt{3} \left( \frac{\sqrt{3}}{2} {\bf n}_x +
\frac{1}{2} {\bf n}_y
\right) \nonumber \\
{\bf a}_2 &= a \sqrt{3} \left( \frac{\sqrt{3}}{2} {\bf n}_x -
\frac{1}{2} {\bf n}_y \right)
\end{align}
where $a \approx 1.4$\AA\ denotes the bond length.
Both the vector and the scalar potential can be expressed in terms
of the strain  tensor $u_{ij}$. The scalar potential is determined
by the trace of the strain tensor, $\phi({\bf r},t)=g_D
(u_{xx}+u_{yy})$, where $g_D$ is estimated to be of order 20-30eV in
Ref.\ \onlinecite{SA02b}. The form of the vector potential is
essentially fixed by symmetry to be\cite{M07,MO08}
\begin{align}
{\bf A}({\bf r}) =  \frac{\hbar\beta}{2a} \left( \begin{array}{c}
2 u_{xy} \\  u_{xx} - u_{yy} \end{array} \right)
\label{gauge}
\end{align}
where $\beta = \partial \log ( t ) / \partial \log( a ) \approx 2-3$. Strictly speaking,
this expression has a small uncertainty in the prefactor since the
theory of elasticity may not accurately describe the displacements
within the unit cell.

In nanotubes, the natural coordinate system, defined by the nanotube
axis, is  rotated by an angle $\theta$ with respect to the
coordinate axes of graphene defined above. Choosing the CNT axis as
the $x$-axis and the direction around the tube as the $y$-axis, the
angle $\theta$ is given by $\cot\theta=\sqrt{3}\frac{n-m}{n+m}$ for
(n,m) carbon nanotubes. It takes the value $\theta=0$
($\theta=\pi/2$) for zigzag (armchair) CNT. The vector potential
takes the form
\begin{align}
{\bf A}({\bf r}) =  \frac{\hbar\beta}{2a}\, D(3\theta)\left( \begin{array}{c}
2 u_{xy} \\ u_{xx} - u_{yy} \end{array} \right),
\label{gauge_CNT}
\end{align}
where $D(3\theta)$ is a rotation matrix. Note that in this equation,
also the  strain tensor is given in the rotated coordinate system.

The acoustic phonon modes of carbon nanotubes and graphene can be
described within  standard elasticity theory. For graphene, the
elastic Lagrangian density for the strain tensor $u_{ij}$ and the
out-of-plane displacement $h({\bf r},t)$ takes the form
\begin{equation}
\label{elastic}
    {\cal L} = T-V_{{\rm stretch}}-V_{{\rm bend}}
\end{equation}
with
\begin{eqnarray}
\label{StretchBend}
&& T = \frac{\rho_0}{2} \left(\dot{\bf u}^2+\dot h^2\right)\nonumber \\
&& V_{{\rm stretch}}^{} = \mu u^2_{ij}+\frac{1}{2} \lambda u^2_{kk} \\
&& V_{{\rm bend}}^{} = \frac{1}{2}\kappa \left(\nabla^2 h\right)^2\nonumber
\end{eqnarray}
in terms of the 2D mass density $\rho_0$, the Lam\'e coefficients
$\mu$ and $\lambda$  characterizing the in-plane rigidity of the
lattice, and the bending rigidity $\kappa$. The same Lagrangian also
applies to carbon nanotubes of radius $R$, when replacing the
bending energy by
\begin{eqnarray}
V_{{\rm bend}} = \frac{1}{2}\kappa \left(\nabla^2 h+\frac{h}{R^{2}}\right)^2.
\end{eqnarray}
The strain tensor takes the form $u_{ij}=(1/2)[\partial_i
u_{j}+\partial_j  u_{i}+(\partial_i h)(\partial_j h)]$ for graphene and
\begin{align}
u_{xx} &= \frac{\partial u_x}{\partial x} \nonumber \\
u_{yy} &= \frac{\partial u_y}{\partial y} + \frac{h}{R} \nonumber \\
u_{xy} &= \frac{1}{2} \left( \frac{\partial u_y}{\partial
x} + \frac{\partial u_x}{\partial y} \right),
\label{tensor_CNT}
\end{align}
for carbon nanotubes. Here, ${\bf u}$ denotes the displacements within the graphene sheet
and $h$  the displacement in the perpendicular (for CNT: radial) direction.

\subsection{Phonon damping}
\label{PhononDamping}

As argued in the Introduction, there is a close relation between
phonon  damping and the conductivity tensor. This relation can be
obtained formally by computing the shift $\Delta\omega$ in the
phonon frequency due the electron-phonon coupling. The damping rate
is then given by $\Gamma = 2{\rm Im} \Delta \omega$. Consider first
the case in which the phonon is associated with a vector potential
${\bf A}({\bf r},t)$. We can express the vector potential in terms
of phonon creation and annihilation operators, $b_{\bf q}$ and
$b^\dagger_{\bf q}$, respectively,
\begin{eqnarray}
  {\bf A}({\bf r},t) &=&\frac{1}{2\Omega} \sum_{\bf q}{\cal A}_{\bf q} \exp(i{\bf qr})
  \nonumber\\   &\times&
  \left( b_{\bf q}\exp(-i\omega_{\bf q}t) + b^\dagger_{-\bf q}\exp(i\omega_{\bf q}t) \right)
\end{eqnarray}
with
\begin{eqnarray}
  {\cal A}_{\bf q} = \frac{\hbar\beta}{2a} \sqrt{ \frac{2\hbar\Omega}{ \rho_0 \omega_{\bf q}}} [D(3\theta)M_{\bf q}{\bf
  \hat p}] .
\end{eqnarray}
Here, we defined the matrix
\begin{equation}
  M_{\bf q} = \left(\begin{array}{ccc}
  iq_y & iq_x & 0 \\
  iq_x & -iq_y & -1/R
  \end{array}\right)
\end{equation}
and a unit vector ${\bf \hat p}$ describing the mode polarization in
$u_x, u_y$, and $h$-direction, which takes the form ${\bf\hat
p}=(1,0,i\lambda q_x R/(2\mu+\lambda))$ for the longitudinal
stretching mode, ${\bf\hat p}=(0,1,0)$ for the twist mode, and
${\bf\hat p}=(0,0,1)$ for the radial breathing mode [as  obtained
from the Euler-Lagrange equations for the elastic Lagrangian Eq.\
(\ref{elastic})]. The surface area of the nanotube is denoted by
$\Omega=(2\pi R) L$ and the mode dispersion by $\omega_{\bf q}$. The
corresponding results for the acoustic modes of graphene follow by
taking $R\to \infty$ and setting $\theta=0$. (The flexural modes of
of graphene will be discussed separately below.)

In second order perturbation theory, we then find for the  damping
rate of a phonon with wavevector ${\bf q}$,
\begin{eqnarray}
  \Gamma_{\bf q} &=& \frac{\pi}{2\hbar \Omega^2} \sum_{\epsilon_\beta > E_F} \sum_{\epsilon_\alpha<E_F}| \langle \beta|v_F{\bf\sigma}\cdot
  {\cal A}_{\bf q}\exp(i{\bf qr})|\alpha\rangle|^2 \nonumber\\
  &&\times \delta(\epsilon_\beta - \epsilon_\alpha -\hbar\omega_{\bf q})
\end{eqnarray}
Using the Kubo formula, we can express $\Gamma_{\bf q}$ in terms of
the dissipative (real and symmetric) contribution $\sigma_{s}({\bf
q},\omega)$ to the conductivity tensor,
\begin{eqnarray}
   \sigma_{s;kl}({\bf q},\omega) &=& \frac{\pi e^2}{\Omega\omega}\,
   {\rm Re}\sum_{\epsilon_\beta > E_F} \sum_{\epsilon_\alpha<E_F}    \langle \beta|v_k\exp(i{\bf qr})|\alpha\rangle  \nonumber\\
  &\times& \langle \alpha|v_l\exp(-i{\bf qr})|\beta\rangle \delta(\epsilon_\beta - \epsilon_\alpha -\hbar\omega).
\label{Kubo}
\end{eqnarray}
Here, ${\bf v}=v_F {\bf \sigma}$ denotes the velocity operator.
Note that we employ the two-dimensional conductivity unless
explicitly stated otherwise. Thus, we can express the damping rate
as
\begin{equation}
  \Gamma_{\bf q} = \frac{\omega_{\bf q}}{2e^2\hbar \Omega} \sigma_{s;ij}({\bf q},\omega)
  ({\cal A}^*_{{\bf q};i}+\frac{q_i}{\omega_{\bf q}}\phi^*_{\bf q})({\cal A}_{{\bf q};j}+\frac{q_j}{\omega_{\bf q}}\phi_{\bf q}).
  \label{damping}
\end{equation}
Here, we have also included the effect of the scalar potential $\phi$ and defined
\begin{equation}
  \phi_{\bf q} = \frac{g_D}{1+v({\bf q})\Pi({\bf q},\omega)}
  \sqrt{\frac{2\hbar\Omega}{\rho_0\omega_{\bf q}}} {\bf M}_{\bf q}^\phi\cdot {\bf \hat p}
  \label{ScalarPotential}
\end{equation}
in terms of the vector
\begin{equation}
  {\bf M}_{\bf q}^\phi =(iq_x,iq_y,1/R).
\end{equation}
Note that we have included the effects of screening in the
contribution originating from the scalar potential. ($v({\bf q})$
denotes the Coulomb interaction and $\Pi({\bf q},\omega)$ the
polarization operator.)

\subsection{Basic physical picture: Damping of the radial breathing mode in armchair CNT}
\label{Picture}

Before we embark on a systematic investigation of phonon damping
based on the expressions derived above, we would like to illustrate
the basic physics in the context of the radial breathing mode of
armchair carbon nanotubes. We will do this in the context of a
semiclassical approach which clearly brings out the physics and
complements the more quantum mechanical approach taken in the
remainder of the paper. This example also shows that our approach
yields damping rates which are of the order of those observed in
experiment.\cite{LLKD04}

For the radial breathing mode of carbon nanotubes, only the radial
displacement $h$ is  nonzero (i.e., ${\bf u}=0$) in the long-wavelength limit. According to Eq.\
(\ref{tensor_CNT}), the corresponding strain tensor takes the form
$u_{xx}=u_{xy}=0$ while $u_{yy}=h/R$. Using that for armchair carbon
nanotubes $\theta=\pi/2$, the synthetic vector potential takes the
form
\begin{equation}
{\bf A}({\bf r},t) = \frac{\hbar\beta}{2a} \left(\begin{array}{c} h({\bf r},t)/R \\ 0 \end{array} \right) .
\end{equation}
Similarly, we find for the scalar deformation potential $\phi({\bf
r},t) = g_D h({\bf r},t) / R$.  In the long wavelength limit $q\to
0$, the electric field $-\nabla \phi$ associated with the scalar
potential vanishes. At the same time, the valley electric field
$e{\bf E}=-\partial {\bf A}/\partial t$ originating from the vector
potential remains finite because the RBM dispersion tends to a
finite frequency $\omega_B= [(2\mu+\lambda)/\rho_0 R^2]^{1/2}$ for
$q\to 0$. Thus, the synthetic vector potential gives the dominant
contribution to the damping of the radial breathing mode.  Note also
that the electric field is pointing in the direction along the
nanotube axis.

For clean armchair carbon nanotubes, the damping of the radial
breathing mode vanishes  to leading order,\cite{Setal08b} since the
contribution of the synthetic vector potential to the Hamiltonian is
proportional to $\sigma_x$ and hence commutes with the unperturbed
Hamiltonian. However, since armchair carbon nanotubes are metallic,
there will be damping in the presence of disorder. Describing the
optical conductivity of carbon nanotubes within a Drude model, we
have
\begin{equation}
  \sigma(\omega)=\frac{\sigma_{dc}}{1-i\omega \tau}
\end{equation}
in terms of the mean free time $\tau$. From the Einstein relation,
the (one-dimensional) {\it dc} conductivity  takes the form
$\sigma_{dc}=\frac{Ne^2}{\pi\hbar}\ell$, where $\ell=v_F\tau$
denotes the electronic mean free path and $N$ counts the spin and
valley degeneracy. We can now compute the damping rate of the radial
breathing mode, $\Gamma$, by equating the time derivative of the
elastic energy $\cal E$ with the Joule heating associated with ${\rm Re}
\sigma(\omega)$.

The elastic energy of the radial breathing mode (per unit length)
can be readily obtained from the elastic Lagrangian so that we find
\begin{align}
  \frac{d{\cal E}}{d t} &=
  \frac{d}{dt}\left[(2\pi R)\rho_0 \omega_B^2 |h(\omega_B)|^2 \right] \nonumber \\
  &=-{\rm Re}\sigma(\omega_B)|{\bf E}(\omega_B)|^2 \nonumber \\
 &  = \hbar \omega_B \Gamma
  \label{energy_diss}
\end{align}
Inserting $e|{\bf E}(\omega)|=\frac{\hbar\beta}{2a}\frac{\omega_B}{R}|h(\omega)|$ for  the
synthetic electric field, we find a damping rate $\Gamma$ of
\begin{equation}
 \Gamma = \frac{N\hbar\beta^2}{8\pi^2(\rho_0 a^2)R^3}\frac{\ell}{1+(\omega_B\tau)^2}.
 \label{width_B}
\end{equation}
An interesting aspect  of this expression is that the damping rate
falls off with the third power of the nanotube radius. This may be
useful guidance to enter into the regime of current-driven
nonequilibrium in nanoelectromechanical devices.

Numerical estimates of Eq.\ (\ref{width_B}) yield relaxation times
of the order of nanoseconds for an elastic mean free path of 1
$\mu$m and a diameter of order 1nm. This is of the same magnitude as the relaxation times observed
in experiment.\cite{LLKD04}

\section{Disorder effects on phonon damping}
\label{disorder}

\subsection{Carbon nanotubes}
\label{CNT}

Due to the weakness of screening in one dimension, the scalar
potential  remains relevant in carbon nanotubes. Indeed, one readily
estimates that in the absence of screening, the ratio of the
electric fields due to vector and scalar potential is of the order
\begin{equation}
  \frac{E_A}{E_\phi}\sim \frac{\omega A}{q\phi}\sim \frac{\hbar c \beta}{a g_D}\sim 10^{-2}
\end{equation}
for phonon modes with a linear dispersion $\omega = cq$. At the same
time,  the suppression of the scalar potential by screening
\begin{equation}
   [1+v({\bf q})\Pi({\bf q},\omega)]^{-1} \simeq [1+ \frac{e^2\nu}{\pi}\ln (1/qR)]^{-1}
\end{equation}
involves, for realistic values of $q$, a factor smaller than but still of order one. Here, $\nu$
denotes the electronic density of states and $R$ the radius of the
nanotube. As a result, we conclude that the damping of the
longitudinal stretching mode is dominated by the effects of the
scalar potential. In contrast, the damping of the
radial breathing mode (whose dispersion approaches a finite
frequency as $q\to 0$) will be dominated by the vector potential. In
the following, we will only discuss the dominant dissipation channel
for the various phonon modes of carbon nanotubes. An evaluation of
the subdominant dissipation channel would, however, be
straight-forward.

\subsubsection{Radial breathing mode}

In order to make contact with the literature,\cite{LLKD04}  it is
instructive to start with the damping of the radial breathing mode
in clean carbon nanotubes. One readily establishes that only the
$y$-component of the vector potential
\begin{equation}
  {\cal A}_{\bf q} = -\frac{\hbar\beta}{2aR} \sqrt{\frac{2\hbar\Omega}{\rho_0\omega_{\bf q}}}\left(
  \begin{array}{c} -\sin(3\theta) \\ \cos(3\theta) \end{array}\right)
\label{RBM-A}
\end{equation}
contributes to dissipation. Evaluating Eq.\ (\ref{Kubo}), we obtain
for the corresponding component of the dissipative conductivity
tensor
\begin{equation}
  \sigma_{yy}(\omega) = \frac{Ne^2v_F}{2\hbar R\omega}\theta(\omega-2v_Fk_F).
\label{RBM-sigma}
\end{equation}
Inserting Eqs.\ (\ref{RBM-A}) and (\ref{RBM-sigma}) into the general
expression Eq.\ (\ref{damping}), we recover for the damping rate of
the RBM of clean carbon nanotubes\cite{LLKD04}
\begin{equation}
  \Gamma_{\rm RBM} = \frac{N\hbar\beta^2 v_F}{8(\rho_0a^2)R^3 \omega}\cos^2(3\theta) \theta(\omega-2v_Fk_F).
\end{equation}
Specifically, this expression confirms the absence of  damping for
armchair CNT where $\theta=\pi/2$. This shows that our approach is
equivalent to standard approaches to damping of phonon modes.

The strength of making the relation with the synthetic  electric
fields lies in allowing us to go beyond the limit of clean and
non-interacting samples. Including disorder for the armchair CNT,
there will also be damping due to electric fields pointing along the
CNT. Again using a Drude expression for the frequency-dependent
conductivity, one recovers from Eq.\ (\ref{damping}) the
semiclassical result given above in Eq.\ (\ref{width_B}).

\begin{figure}[!t]
\begin{center}
\includegraphics[width=6cm,angle=0]{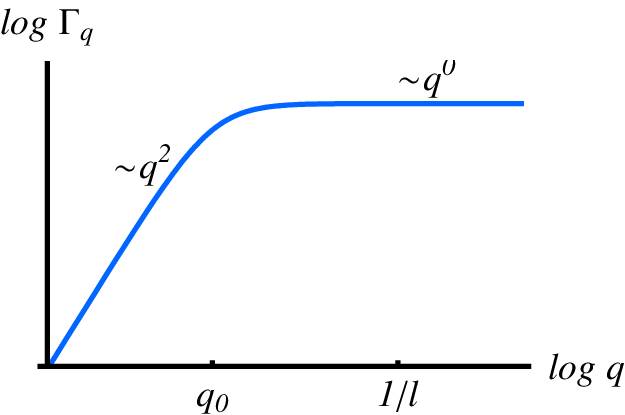}
\caption[fig]{(Color online). Sketch of the damping rate as function of wavevector for the longitudinal
stretching mode of carbon nanotubes.} \label{long_CNT}
\end{center}
\end{figure}

\subsubsection{Longitudinal stretching mode}

At long wavelengths, the longitudinal stretching mode of carbon
nanotubes  has a much lower frequency $\omega=cq$ than the radial
breathing mode and we can make a diffusive ansatz for the dynamic
conductivity,
\begin{equation}
  \sigma({\bf q},\omega) = \frac{-i\omega \sigma_{\rm dc}}{-i\omega + Dq^2},
\end{equation}
which is valid when $q\ell\ll 1$ and $D = v_F \ell / 2$ is the
diffusion coefficient. Note that the diffusion pole in this
expression is protected by charge conservation since the scalar
potential drives ordinary charge currents. The dissipative
conductivity is then given by
\begin{equation}
 {\rm Re}\sigma({\bf q},\omega) = \frac{\omega^2\sigma_{\rm dc}}{\omega^2+(Dq^2)^2} .
\end{equation}
The precise $q$ dependence of the conductivity depends sensitively
on the  relative magnitudes of $\omega=cq$ and $Dq^2$. To clearly
bring out the $q$ dependence of the damping rate, we consider the
long-wavelength regime $q\ll q_0$, dominated by $\omega$, and the
short-wavelength regime $q\gg q_0$, dominated by $Dq^2$, separately.
Here, the characteristic wavevector dividing between these two
regimes is given by
\begin{equation}
  q_0 = 2\frac{c}{v_F}\frac{1}{\ell}.
\end{equation}
Evaluating the scalar potential Eq.\ (\ref{ScalarPotential}) for the
longitudinal stretching mode and inserting the resulting expression
into Eq.\ (\ref{damping}), we obtain the damping rate
\begin{equation}
\Gamma_{\bf q} \simeq \frac{2N{g_D^*}^2 }{\pi^2\hbar\rho_0 R v_F^2\ell}\left\{ \begin{array}{ccc}
(q/q_0)^2 & & q\ll q_0 \\
1 & & q\gg q_0 \end{array}
\right. .
\label{LSM-damping}
\end{equation}
Here, we defined the renormalized coupling constant
\begin{equation}
  g_D^* = \frac{g_D}{1+\frac{Ne^2}{\pi^2\hbar v_F}\ln(1/qR)},
\end{equation}
which, strictly speaking, still includes a weak logarithmic $q$ dependence.

Eq.\ (\ref{LSM-damping}) predicts that the damping rate increases
quadratically with $q$ for long wavelengths and saturates to a ${\bf
q}$-independent constant for shorter wavelengths $q\gg q_0$. This
behavior is sketched in Fig.\ \ref{long_CNT}. It is interesting to compare
$\Gamma_{\bf q}$ to the mode frequency $\omega_{\bf q}$ in order to
see whether the longitudinal stretching mode can become overdamped
for some region of wavevectors. Clearly, the ratio $\Gamma_{\bf
q}/\omega_{\bf q}$ is maximal for $q=q_0$. Remarkably, the maximal
value of this ratio becomes independent of the elastic mean free
path and thus quite universal,
\begin{equation}
  \frac{\Gamma_{{\bf q}_0}}{\omega_{{\bf q}_0}} = \frac{N{g_D^*}^2}{\pi^2\rho_0 R \hbar v_F c^2}.
\end{equation}
Inserting numbers typical of carbon nanotubes into this expression,
one finds values of order $0.1$ showing that even though the
longitudinal stretching mode remains underdamped at all wavelengths,
damping can be quite significant.

\begin{figure}[!t]
\begin{center}
\includegraphics[width=6cm,angle=0]{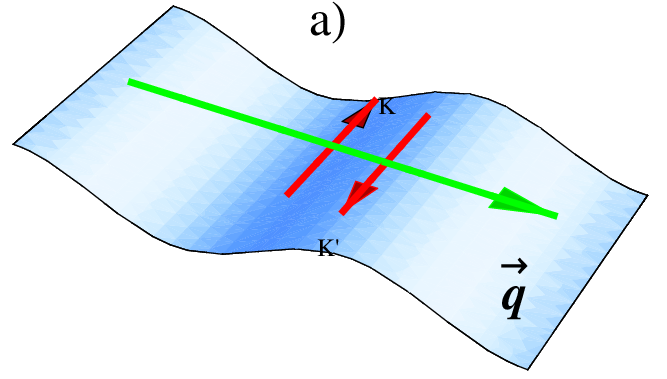}
\includegraphics[width=6cm,angle=0]{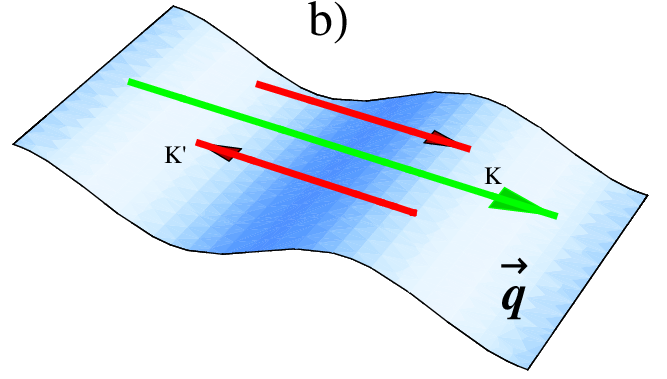}
\caption[fig]{(Color online) Orientation of the synthetic  electric
field at the two K-points of the dispersion (red arrows) with
respect to the phonon wavevector ${\bf q}$ for (a) the purely
transverse and (b) the purely longitudinal situation. As shown in
Eqs.\ (\ref{polarization1}) - (\ref{polarization2}), the synthetic
electric field is in general neither purely parallel nor
transverse.} \label{long_trans}
\end{center}
\end{figure}

\subsection{Graphene}

\subsubsection{Longitudinal and transverse acoustic (in-plane) phonons}

The analysis of the previous section can be extended to phonon
modes of graphene. However, phonon damping in (doped) graphene will
typically be dominated by the vector potential. At long wavelengths,
this is a consequence of the much stronger screening in two
dimensions than in the carbon nanotube setting,
\begin{equation}
  \frac{1}{1+v({\bf q})\Pi({\bf q},\omega)} \simeq \frac{q}{q+q_{\rm TF}},
\end{equation}
where $q_{\rm TF}=2\pi e^2\nu $ denotes the Thomas-Fermi  wavevector
of graphene. (For a density of approximately $10^{12}$cm$^{-2}$, the
screening length is of the order of tens of nm.) At short phonon
wavelengths (relative to the elastic mean free path), this emerges
from the fact that unlike $-\nabla\phi$, ${\bf E}=-\partial{\bf
A}/\partial t$ will almost always have a significant {\it
transverse} component (relative to the wavevector ${\bf q}$, cp.\
Fig.\ \ref{long_trans}). \footnote{We note that the description of
the synthetic fields as electric and magnetic fields is motivated by
their effect on the Dirac carriers. At the same time, the dynamics
of these fields is entirely controlled by elasticity theory which
has little resemblence to electrodynamics.} It turns out that in
this range of wavevectors, the transverse conductivity (and hence
the associated damping) is much larger than the longitudinal
conductivity. In addition, even when the synthetic electric field is
purely longitudinal, the weaker bare coupling of the vector
potential is partially offset by the fact that dissipation by
longitudinal valley currents is very sensitive to (and strongly
enhanced by) disorder-induced intervalley scattering. (It is
interesting to note that in the context of multi-valley
semiconductors, the importance of intervalley scattering in acoustic
attenuation is actually known for many decades.\cite{Pomerantz62})

Using Eq.\ (\ref{gauge}), we can readily obtain the longitudinal
and transverse components of the vector potential. Consistent with
the lattice symmetry, we find
\begin{eqnarray}
  {\cal A}_{{\bf q},\parallel}&=&\frac{\hbar\beta}{2a}\sqrt{\frac{2\hbar\Omega}{\rho_0\omega_{\bf q}}}iq\sin3\theta_{\bf q} \label{polarization1}  \\
  {\cal A}_{{\bf q},\perp}&=&\frac{\hbar\beta}{2a}\sqrt{\frac{2\hbar\Omega}{\rho_0\omega_{\bf q}}}iq\cos3\theta_{\bf q}
\end{eqnarray}
for the longitudinal phonon and
\begin{eqnarray}
  {\cal A}_{{\bf q},\parallel}&=&\frac{\hbar\beta}{2a}\sqrt{\frac{2\hbar\Omega}{\rho_0\omega_{\bf q}}}iq\cos3\theta_{\bf q} \\
  {\cal A}_{{\bf q},\perp}&=&\frac{\hbar\beta}{2a}\sqrt{\frac{2\hbar\Omega}{\rho_0\omega_{\bf q}}}iq\sin3\theta_{\bf q}
\label{polarization2}
\end{eqnarray}
for the transverse phonons. Here, $\theta_{\bf q}$ denotes the angle
between  the direction of the wavevector ${\bf q}$ and the $x$-axis.

At finite ${\bf q}$, the dissipative conductivity becomes a
symmetric tensor which is diagonal in a coordinate system whose axes
are parallel and perpendicular to the wavevector ${\bf q}$,
\begin{equation}
  \sigma({\bf q},\omega) = \left(\begin{array}{cc} \sigma_\parallel({\bf q},\omega) & 0 \\
  0 & \sigma_\perp({\bf q},\omega) \end{array}\right).
\end{equation}
For doped graphene in the diffusive regime $q\ell\ll 1$, the
longitudinal and transverse conductivities can be obtained from
hydrodynamic equations for charge densities and currents which
follow from a Boltzmann equation. One finds the continuity equations
\begin{eqnarray}
\frac{\partial n_1}{\partial t}+\nabla\cdot{\bf j}_1 &=& -\frac{1}{\tau_V}(n_1-n_2) \\
\frac{\partial n_2}{\partial t}+\nabla\cdot{\bf j}_2 &=& -\frac{1}{\tau_V}(n_2-n_1)
\end{eqnarray}
as well as Ohm's laws
\begin{eqnarray}
\left(\frac{1}{\tau}+\frac{1}{\tau_V}\right){\bf j}_1 - \frac{1}{\tau_V}{\bf j}_2 &=&
 -\frac{v_F^2}{2}{\bf \nabla}\rho_1 +\frac{e^2\nu v_F^2}{2}{\bf E}_1 \nonumber \\
 \label{hydro-disorder1}\\
- \frac{1}{\tau_V}{\bf j}_1+ \left(\frac{1}{\tau}+\frac{1}{\tau_V}\right){\bf j}_2
 &=& -\frac{v_F^2}{2}{\bf \nabla}\rho_2 +\frac{e^2\nu v_F^2}{2}{\bf E}_2 \nonumber\\
 \label{hydro-disorder2}
\end{eqnarray}
Here, the indices 1 and 2 label the valleys, $\nu$ denotes the
density of  states, and $1/\tau$ and  $1/\tau_V$ are the
intra-valley and inter-valley scattering rates due to disorder,
respectively. We left out the effects of induced electric fields
from these equations because screening does not affect the valley
odd channel. We can now obtain the valley-odd conductivity relating
${\bf E}_-={\bf E}_1 - {\bf E}_2$ to ${\bf j}_-={\bf j}_1 - {\bf
j}_2$ by taking the difference between the two equations
(\ref{hydro-disorder1}) and (\ref{hydro-disorder2}), combined with
the continuity equation in the valley-odd channel.

\begin{figure}[!t]
\begin{center}
\includegraphics[width=4cm,angle=0]{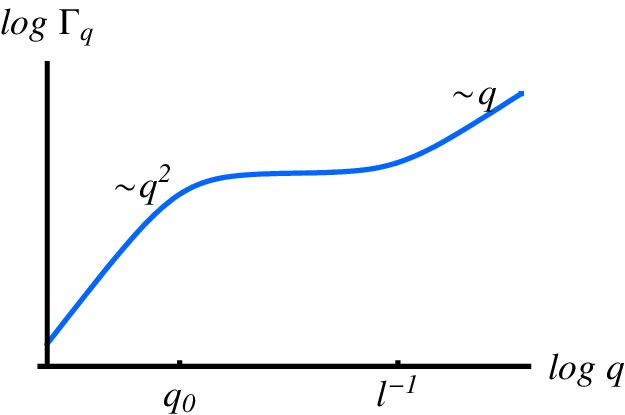}
\includegraphics[width=4cm,angle=0]{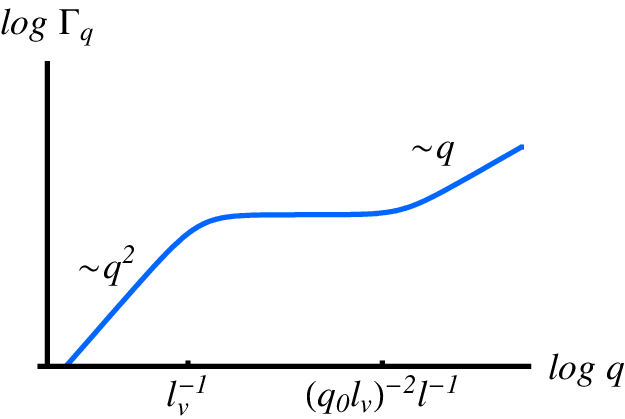}
\caption[fig]{(Color online). Schematic dependence of  the damping
rate $\Gamma_{\bf q}$  on wavevector ${\bf q}$ of acoustic phonons
in graphene when the synthetic electric field is purely
longitudinal, cp.\ Eqs.\ (\ref{Gamma_long1}) and
(\ref{Gamma_long2}). This applies when the wavevector points in a
narrow cone around the zigag (armchair) direction for the
longitudinal (transverse) acoustic phonon. Left: $q_0 >
\ell_V^{-1}$. Right: $q_0 < \ell_V^{-1}$.} \label{zero_temp_long}
\end{center}
\end{figure}

For the longitudinal conductivity, we find
\begin{equation}
  \sigma_\parallel({\bf q},\omega) = \frac{(-i\omega+\frac{2}{\tau_V}) \sigma_{\rm dc}}{-i\omega + Dq^2+\frac{2}{\tau_V}}.
\label{sigma-graphene}
\end{equation}
The reason for the cutoff of the diffusion pole by the intervalley
scattering rate $1/\tau_V$ is that there is no conservation law
associated with valley currents which is analogous to charge
conservation for ordinary charge currents. Despite the weakness of
intervalley scattering (originating from atomic-scale defects) compared to intravalley scattering, it is
actually rather important to include $1/\tau_V$. From Eq.\
(\ref{sigma-graphene}), the dissipative conductivity takes the form
\begin{equation}
  {\rm Re}\, \sigma_\parallel({\bf q},\omega) = \frac{[\omega^2+
  \frac{2}{\tau_V}(Dq^2+\frac{2}{\tau_V})] \sigma_{\rm dc}}{\omega^2 + (Dq^2+\frac{2}{\tau_V})^2}.
  \label{ReSigmaLong}
\end{equation}
While it is evidently possible to work with this complete
expression, it  is more instructive to analyze the various limiting
cases.  Except for the scale $q_0=(2c/v_F)(1/\ell)$ introduced
above, this expression involves the intervalley scattering length
$\ell_V=[\frac{1}{2}D\tau_V]^{1/2}$ as a second length scale. When
$q\ell_V\ll 1$, one finds that ${\rm Re}\,\sigma_\parallel({\bf
q},\omega) \simeq \sigma_{\rm dc}$. In the opposite limit
$q\ell_V\gg 1$, one obtains
\begin{equation}
   {\rm Re}\,\sigma_\parallel({\bf q},\omega) \simeq \left\{ \begin{array}{ccc} \sigma_{\rm dc} & & q\ll q_0 \\
   \frac{1+(q_0\ell_V)^2}{(q\ell_V)^2}\sigma_{\rm dc} & & q\gg q_0 \end{array} \right.
\end{equation}
Transverse fields do not induce any charge densities and therefore, we find
\begin{equation}
   \sigma_\perp({\bf q},\omega) = \sigma_{\rm dc}
   \label{ReSigmaTrans}
\end{equation}
for the transverse conductivity at any $q\ll 1/\ell$.

For doped graphene in the ballistic regime $q\ell\gg 1$, we can
obtain the dynamic conductivity $\sigma({\bf q},\omega)$ from the
Boltzmann equation. Heuristically, we can obtain $\sigma({\bf
q},\omega)$ (up to numerical prefactors) by making the replacements
$Dq^2\to v_F q$ and $1/\ell \to q$ in the diffusive results Eqs.\
(\ref{ReSigmaLong}) and (\ref{ReSigmaTrans}). A more formal
derivation is relegated to the Appendix. For transverse electric
fields, disorder-induced intervalley scattering is irrelevant (since
no valley charge densities are induced) so that the transverse
conductivity becomes
\begin{eqnarray}
  \sigma_\perp({\bf q},\omega) = \frac{Ne^2}{h}\frac{k_F}{q}
\end{eqnarray}
for $\omega\ll v_F q$. It is interesting to note that this
transverse  conductivity $\sigma_\perp$ also controls the
attenuation of surface acoustic waves in the fractional quantum Hall
effect at Landau level filling factor $\nu=1/2$.\cite{Halperin} In
contrast, intervalley scattering is important for the longitudinal
conductivity when $q_0\ell_V\ll 1$,
\begin{eqnarray}
  \sigma_\parallel({\bf q},\omega) = \left\{ \begin{array}{ccc} \frac{e^2\nu}{q^2\tau_V} & & q\ll
  \frac{1}{(q_0\ell_V)^2}\frac{1}{\ell}\\
  \frac{ e^2 \nu\omega^2}{v_Fq^3} & & q\gg \frac{1}{(q_0\ell_V)^2}\frac{1}{\ell}\end{array}\right.
  \label{LongCond}
\end{eqnarray}
while for $q_0\ell_V\gg 1$, we find
\begin{eqnarray}
  \sigma_\parallel({\bf q},\omega) =  \frac{ e^2 \nu\omega^2}{v_Fq^3}
\end{eqnarray}
for all $q\gg 1/\ell$. One readily establishes that
$\sigma_\parallel/\sigma_\perp \sim (c/v_F)^2\sim 10^{-4}$ for sufficiently
large $q$,  confirming that the damping will be dominated by the
transverse conductivity except in a very narrow range of directions
of ${\bf q}$ where the synthetic electric field is almost purely
longitudinal.

\begin{figure}[!t]
\begin{center}
\includegraphics[width=6cm,angle=0]{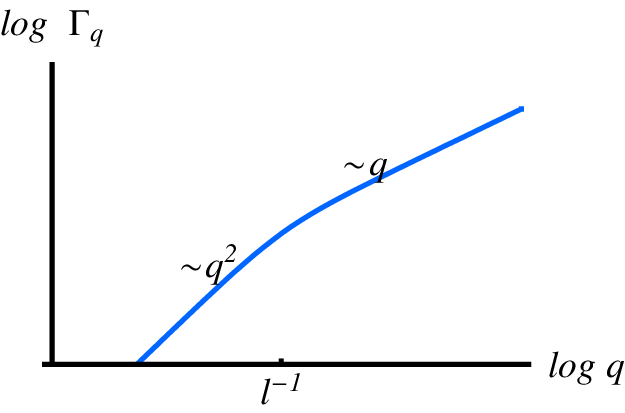}
\caption[fig]{(Color online). Schematic dependence  of the damping
rate $\Gamma_{\bf q}$ on wavevector $q$ for acoustic phonons in graphene when the
tranverse synthetic electric field is dominant, cp.\ Eq.\ (\ref{Gamma_trans1}). This applies for all
directions of ${\bf q}$, except when the wavevector points in a
narrow cone around the zigag (armchair) direction for the
longitudinal (transverse) acoustic phonon.} \label{zero_temp_trans}
\end{center}
\end{figure}

We are now in a position to combine these results with Eq.\
(\ref{damping}) and obtain  the phonon damping rate in graphene over
the full range of wavevectors. Whenever the synthetic electric field
has an appreciable transverse component, we find
\begin{equation}
  \Gamma_{\bf q} = \frac{N\hbar\beta^2 k_F}{8\pi \rho_0 a^2} f(\theta_{\bf q}) \left\{ \begin{array}{ccc}
  q^2\ell/2 & & q\ell \ll 1 \\ q & & q\ell\gg 1
  \end{array}\right. .
  \label{Gamma_trans1}
\end{equation}
Here, we defined the function $f(\theta)$ which is equal to unity
 deep in the diffusive regime, and crosses over to
$f(\theta)=\cos^2 3\theta$ for transverse phonons and
$f(\theta)=\sin^2 3\theta$ for longitudinal phonons\footnote{The
crossover occurs at $q\sim 1/\ell$ when $q_0\ell_V\gg 1$ and at
$q\sim 1/\ell_V$ when $q_0\ell\ll 1$. A more accurate expression for
the function $f(\theta)$ can be readily derived by including both
the longitudinal and the transverse components of the synthetic
electric field in the calculation of the damping rate.}. Thus, we
find that the phonon modes are underdamped at long wavelengths
$q\ell\ll1$, but become marginal for $q\ell\gg1$. Inserting numbers,
we find that $\Gamma_{\bf q}\approx 10^{-2}\omega_{\bf q}$ in this
marginal regime so that the phonon mode remains well defined.

Whenever $f(\theta)$ is close to zero, the damping is dominated by
the longitudinal  conductivity. Specifically, this happens when
${\bf q}$ points in the zigzag (armchair) direction for longitudinal
(transverse) phonons. In these cases, the damping exhibits an
intermediate ${\bf q}$-independent regime in between the quadratic
and the linear wavevector dependence. Specifically, we find
\begin{eqnarray}
  \Gamma_{\bf q} = \frac{N\hbar\beta^2 k_F}{8\pi \rho_0 a^2} \left\{ \begin{array}{ccc}
  q^2\ell/2 & & q\ll q_0 \\ {q_0^2\ell}/{2} & & q_0\ll q \ll \frac{1}{\ell} \\
  ({c}/{v_F})^2q& & q\gg \frac{1}{\ell}
  \end{array}\right.
  \label{Gamma_long1}
\end{eqnarray}
for very weak intervalley scattering $q_0\ell_V\gg 1$ and
\begin{eqnarray}
  \Gamma_{\bf q} = \frac{N\hbar\beta^2 k_F}{8\pi \rho_0 a^2}  \left\{ \begin{array}{ccc}
  q^2\ell/2 & & q\ll \frac{1}{\ell_V} \\ {\ell}/{2\ell_V^2} & & \frac{1}{\ell_V}\ll q \ll \frac{1}{(q_0\ell_V)^2}\frac{1}{\ell} \\
  ({c}/{v_F})^2q& & q\gg \frac{1}{(q_0\ell_V)^2}\frac{1}{\ell}
  \end{array}\right.
  \label{Gamma_long2}
\end{eqnarray}
for stronger intervalley scattering $q_0\ell_V\ll 1$.

Our results for the acoustic (in-plane) phonons of graphene are
summarized in Figs.\ \ref{zero_temp_long} and \ref{zero_temp_trans}.
For transverse synthetic fields, the damping rate crosses over from
a quadratic to a linear dependence on $q$ when $q\sim 1/\ell$. In
contrast, when the damping is dominated by longitudinal fields,
there is an intermediate constant regime for ${\rm
max}\{q_0,1/\ell_V\}\ll q \ll {\rm
max}\{1,1/(q_0\ell_V)^2\}(1/\ell)$. In magnitude, the damping rate
is the same for longitudinal and transverse electric fields for
small $q$ where we find a quadratic dependence on $q$ in both cases.
For larger $q$, the damping is much stronger for transverse fields.
Deep in the ballistic regime $q\ell\gg 1$, where one finds a linear
dependence on $q$ in both cases, the ratio saturates at
approximately $(v_F/c)^2\approx 10^4$.

\subsubsection{Flexural modes}

We now turn to flexural modes of suspended graphene samples which are
characterized by out-of-plane fluctuations $h({\bf r},t)$. A
qualitative difference arises since, by symmetry, $h({\bf r},t)$ appears
quadratically in the strain tensor. As a result, the coupling
between electrons and flexural modes is quadratic rather than
linear. This implies that the dominant damping mechanism involves
both the creation of a particle-hole pair and a lower-energy
flexural phonon.

\begin{figure}[!t]
\begin{center}
\includegraphics[width=8cm,angle=0]{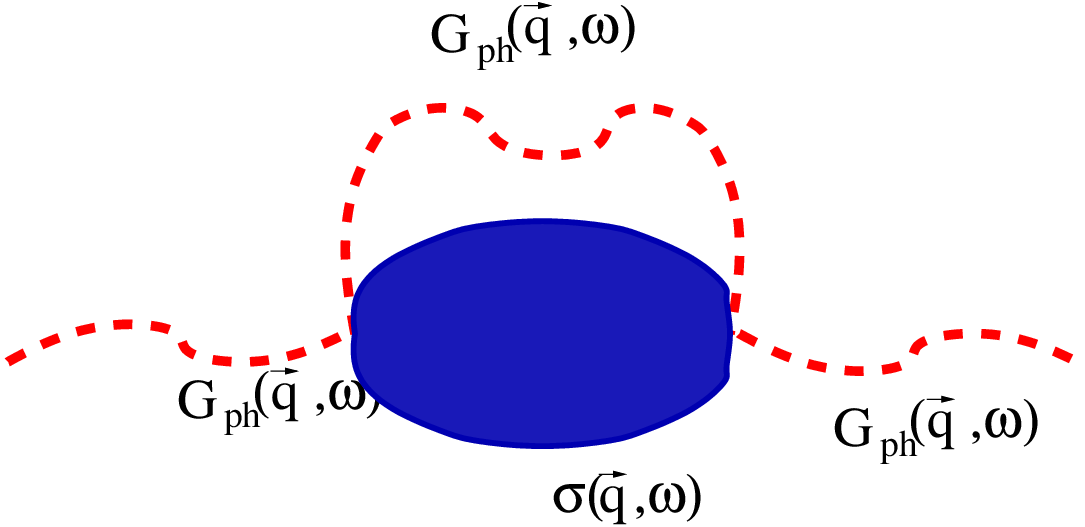}
\caption[fig]{(Color online). Sketch of the Feynman diagram for the
damping rate of flexural phonons.} \label{flexural_fig}
\end{center}
\end{figure}

The corresponding rate can be obtained from Fermi's golden rule  (or
alternatively, the Feynman diagram in Fig.\ \ref{flexural_fig}).
Quantizing the out-of-plane displacements, we find at zero
temperature
\begin{eqnarray}
  \Gamma_{\bf q} &=& \frac{2\pi}{\hbar}\frac{1}{\Omega^4}
    \nonumber \\
     &\times&\sum_{\epsilon_\beta>E_F}\sum_{\alpha<E_F}\sum_{{\bf
  q}^\prime}
  |\langle\beta|v_F{\bf \sigma}\cdot{\cal A}_{{\bf q},{-\bf q}^\prime} e^{i({\bf q}-{\bf q}^\prime)\cdot {\bf
  r}}|\alpha\rangle|^2
  \nonumber\\
  &\times& \delta(\epsilon_\beta - \epsilon_\alpha -\omega_{\bf q} + \omega_{{\bf q}^\prime})
\end{eqnarray}
in terms of
\begin{equation}
  {\cal A}_{{\bf q}_1,{\bf q}_2} = \frac{\hbar\beta}{4a}\left( \begin{array}{c}
  q_{1x}q_{2y} + q_{1y}q_{2x} \\
  q_{1x}q_{2x} - q_{1y}q_{2y}
  \end{array}\right)
    \sqrt{\frac{\hbar\Omega}{2\rho_0\omega_{{\bf q}_1}}}\sqrt{\frac{\hbar\Omega}{2\rho_0\omega_{{\bf q}_2}}}.
\end{equation}
Following the approach of this paper, we rewrite this expression in
terms of the dissipative conductivity Eq.\ (\ref{Kubo}),
\begin{eqnarray}
 \Gamma_{\bf q} &=& \frac{1}{2e^2\hbar\Omega}\left(\frac{\hbar\beta}{2a}\right)^2
 \left(\frac{\hbar}{2\rho_0}\right)^2 \sum_{{\bf q}^\prime} \frac{\omega_{\bf q}-
 \omega_{{\bf q}^\prime}}{\omega_{\bf q}\omega_{{\bf q}^\prime}} \nonumber\\
 &\times&\sigma_{s,ij}({\bf q}-{\bf q}^\prime,\omega_{\bf q}-\omega_{{\bf q}^\prime}) \nonumber\\
 &\times&\left( \begin{array}{c}
  q_{x}q_{y}^\prime + q_{y}q_{x}^\prime \\
  q_{x}q_{x}^\prime - q_{y}q_{y}^\prime
  \end{array}\right)_i
  \left( \begin{array}{c}
  q_{x}q_{y}^\prime + q_{y}q_{x}^\prime \\
  q_{x}q_{x}^\prime - q_{y}q_{y}^\prime
  \end{array}\right)_j
  \label{flexural_damping}
\end{eqnarray}
As we have seen above, the conductivity tensor is dominated by the
transverse  conductivity so that
\begin{equation}
\sigma_{s,ij}({\bf q},\omega) \simeq \sigma_\perp ({\bf q},\omega)\left(\delta_{ij} - \frac{q_iq_j}{q^2}\right).
\end{equation}
At finite temperature $T\gg\omega_{\bf q}$ , the integrand in Eq.\
(\ref{flexural_damping}) is modified to include an extra factor
$T/\hbar\omega_{{\bf q}^\prime}$ which is a consequence of the
existence of a phonon in the final state.

\begin{figure}[!t]
\begin{center}
\includegraphics[width=4cm,angle=0]{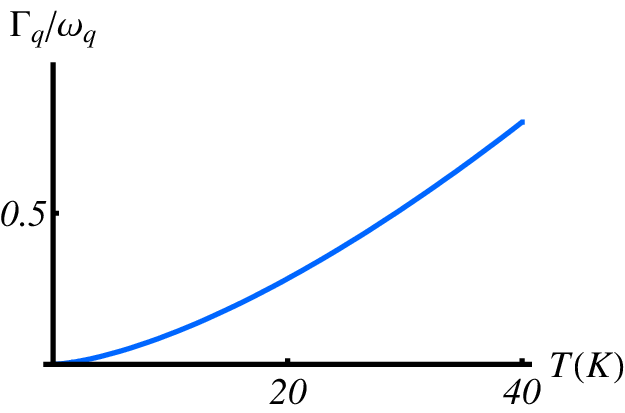}
\includegraphics[width=4cm,angle=0]{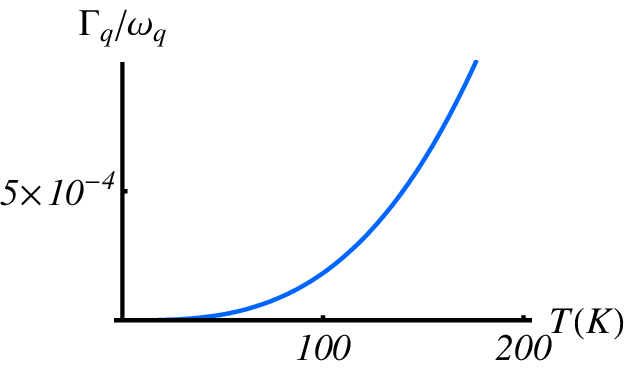}
\caption[fig]{(Color online). Inverse quality factor as function of temperature for long
wavelength flexural vibrations in graphene. The wavevector is $|{\bf
q}|^{-1} = 1\mu$m, the mean free path is $l=100$nm, and the carrier
density is $10^{12}$cm$^{-2}$. Left: No external tension. Right:
External tension $\gamma = 0.02$.} \label{damping_flexural}
\end{center}
\end{figure}

We first focus on ideal graphene membranes for which the dispersion
of flexural  phonons is quadratic, $\omega_{\bf q}=\alpha q^2$ (where $\alpha=\sqrt{\kappa/\rho_0}$). In
this case, we find for the angular average of the damping rate
\begin{equation}
  \Gamma_{\bf q} \sim \left\{\begin{array}{ccc}
  \frac{N\hbar^2 \beta^2 k_F\ell}{\rho_0^2a^2\alpha}q^4 & & q\ell \ll
  1 \\
    \frac{N\hbar^2 \beta^2 k_F}{\rho_0^2a^2\alpha}q^3 & & q\ell \gg 1
   \end{array}\right.
\end{equation}
in the low-temperature limit $T\ll \hbar \alpha q^2$ and
\begin{equation}
  \Gamma_{\bf q} \sim \left\{\begin{array}{ccc}
   \frac{N\hbar^2 \beta^2 k_F\ell}{\rho_0^2a^2\alpha}\left(\frac{T}{\hbar\alpha}\right)^{2}& & T\ll \hbar \alpha/\ell^2 \\
  \frac{N\hbar^2 \beta^2 k_F}{\rho_0^2a^2\alpha}\left(\frac{T}{\hbar\alpha}\right)^{3/2} & & T\gg \hbar \alpha/\ell^2
   \end{array}\right.
\end{equation}
in the high-temperature regime $T \gg \hbar\alpha q^2$. Because of the low frequency
of pure flexural phonons, the vibrations can become overdamped at finite temperature.

If the graphene membrane is under tension $\gamma$  (inducing a term
$\sim(\nabla h)^2$ in the elastic Lagrangian), the dispersion of the
flexural  phonons becomes linear at long wavelengths, $\omega_{\bf
q}=c_f q$. In this case, we find for the angular average of the
damping rate
\begin{equation}
  \Gamma_{\bf q} \sim \left\{\begin{array}{ccc}
  \frac{N\hbar^2 \beta^2 k_F\ell}{\rho_0^2a^2 c_f}q^5   & & q\ell \ll 1\\
    \frac{N\hbar^2 \beta^2 k_F}{\rho_0^2a^2 c_f}q^4  & & q\ell \gg 1
   \end{array}\right.
\end{equation}
in the low-temperature regime $T\ll \hbar c_ f q$ and
\begin{equation}
  \Gamma_{\bf q} \sim \left\{\begin{array}{ccc}
 \frac{N\hbar^2 \beta^2 k_F q \ell}{\rho_0^2a^2 c_f}\left(\frac{T}{\hbar c_f}\right)^4   & & T\ll \hbar c_f/\ell \\
  \frac{N\hbar^2 \beta^2 k_F  q}{\rho_0^2 a^2 c_f}\left(\frac{T}{\hbar c_f}\right)^3 & & T\gg \hbar c_f /\ell
   \end{array}\right.
\end{equation}
in the high-temperature $T\gg \hbar c_f q$ regime.\footnote{In principle, there are also
intermediate cases in  which the dispersion is linear at long
wavelengths due to tension but becomes quadratic at larger $q$. Our
results can be extended to this situation but we refrain from giving
explicit results.}

Actual suspended samples are typically expected to be under  some
degree of tension. The magnitude of phonon damping will then depend
critically on the sample-specific degree of tension, which enters
through the mode velocity $c_f=c\sqrt{\gamma}$. It is interesting to point out that
unlike in-plane modes, the damping of flexural modes is temperature
dependent in the degenerate limit $E_F\gg T$, even in the absence of
electron-electron interactions.

\section{Coulomb interaction effects on phonon damping}
\label{interaction}

\subsection{Damping by valley Coulomb drag in
graphene}\label{subsec:drag}

\begin{figure}[!t]
\begin{center}
\includegraphics[width=4cm,angle=0]{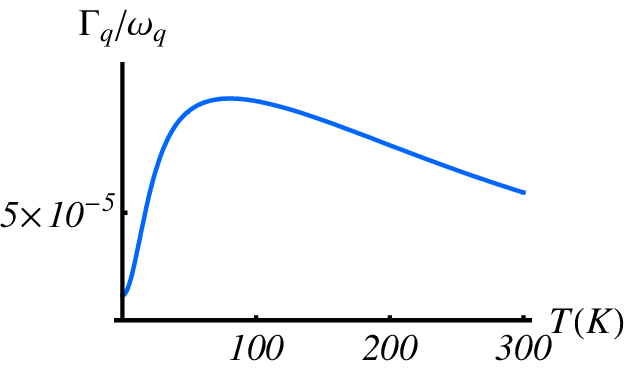}
\includegraphics[width=4cm,angle=0]{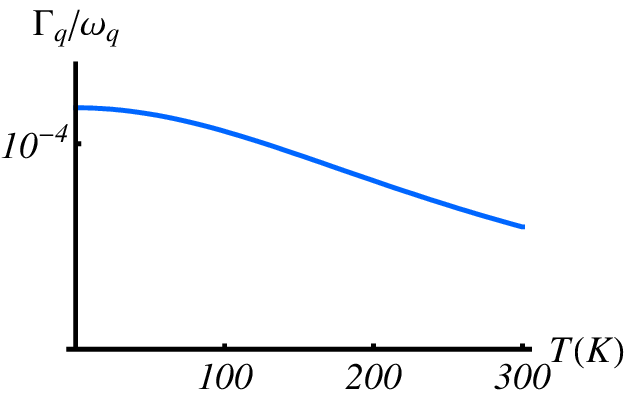}
\caption[fig]{(Color online). Temperature dependence of the  inverse
quality factor  $\Gamma_{\bf q}/\omega_{\bf q}$ for acoustic phonons
in graphene, induced by valley Coulomb drag, when damping  is
dominated by longitudinal (left) and transverse (right) synthetic
electric field. Calculations have been done for an elastic mean free
path $\ell =100$nm, $\tau_V = 200 \tau$, $| {\bf q} | = (1\mu{\rm
m})^{-1}$, and a carrier density $n = 10^{12}$cm$^{-2}$. Note the
enhancement of $\Gamma_{\bf q}$ for transverse synthetic fields.}
\label{finite_temp}
\end{center}
\end{figure}

Electron-electron interactions lead to interesting temperature
dependence of the phonon  damping which persists even in clean
samples. In fact, it is a well-known effect of spintronics that spin
currents decay due to  electron-electron interactions even in the
absence of disorder since unlike charge currents, they are not
protected by momentum conservation. By analogy, the valley currents
driven by the electric fields associated with the vector potential
will dissipate by intervalley Coulomb scattering. A theory of this
effect must go beyond exisiting works on Coulomb drag in that one
has to account for the {\it ac} nature of the driving valley-odd
electric field.

This is easily accomplished in the diffusive  limit where we can
amend the hydrodynamic  equations to include the intervalley
electron-electron scattering. While the continuity equations remain
unchanged, Ohm's laws take the form
\begin{eqnarray}
&&\left(\frac{1}{\tau}+\frac{1}{\tau_V}+
\frac{1}{\tau_D}\right){\bf j}_1 - \left(\frac{1}{\tau_D}+\frac{1}{\tau_V}\right){\bf j}_2 \nonumber\\
 &&\,\,\,\,\,\,\,\,\,\,\,\,\,\,\,\,\,\,\,\,\,\,\,\,\,\,\,\,\,\,=
 -\frac{v_F^2}{2}{\bf \nabla}\rho_1 +\frac{e^2\nu v_F^2}{2}{\bf E}_1 \\
&&- \left(\frac{1}{\tau_D}+\frac{1}{\tau_V}\right) {\bf j}_1+ \left(\frac{1}{\tau}+
\frac{1}{\tau_V}+\frac{1}{\tau_D}\right){\bf j}_2  \nonumber\\
&&\,\,\,\,\,\,\,\,\,\,\,\,\,\,\,\,\,\,\,\,\,\,\,\,\,\,\,\,\,\,=
 -\frac{v_F^2}{2}{\bf \nabla}\rho_2 +\frac{e^2\nu v_F^2}{2}{\bf E}_2
 \label{hydro}
\end{eqnarray}
Here, $1/\tau_D\sim T^2/E_F$ (possibly up to logarithmic  factors
from $2k_F$ scattering) is the intervalley scattering rate
responsible for the drag effect. (Coulomb drag in doped graphene is
expected to have the same features as in other 2D electronic
systems.\cite{THS07}) It is then evident that valley Coulomb drag
can be accounted for by evaluating the $dc$ conductivity
$\sigma_{dc}$ and the diffusion constant $D$ with the effective
scattering rate
\begin{equation}
  \frac{1}{\tau_{\rm eff}} = \frac{1}{\tau} + \frac{2}{\tau_V}+\frac{2}{\tau_D}.
\end{equation}
Thus, the dissipation is dominated by the drag effect (disorder
scattering) when $T>T^*$ ($T<T^*$), where
\begin{equation}
  T^* \sim \sqrt{\frac{E_F}{\tau}} \sim \frac{E_F}{\sqrt{k_F\ell}}.
\end{equation}
Here, we used that for realistic samples, disorder predominantly causes intravalley scattering.
For realistic parameters in suspended graphene, we estimate $T^*\approx 10-100$K.

The inclusion of valley Coulomb drag allows us to discuss the
temperature  dependence of the intrinsic phonon damping in clean
(but doped) graphene. The intervalley electron-electron scattering
leads to a temperature-dependent mean free path $\sim \hbar v_F
E_F/T^2$. Thus, when holding the wavevector ${\bf q}$ fixed, we find
that the damping will follow the behavior of the ballistic
(diffusive) regime for $T<T_{\bf q}^{*}$ ($T>T_{\bf q}^{*}$), where
\begin{equation}
  T_{\bf q}^{*}\sim \sqrt{E_F(\hbar v_F q)}.
\end{equation}
As a result, we find for fixed ${\bf q}$ and when the damping is
dominated by the  transverse synthetic electric field that
$\Gamma_{\bf q}$ is constant for temperatures smaller than $T_{\bf
q}^{*}$ (determined by the transverse conductivity in the ballistic
regime) and {\it decreases} monotonically as $\propto 1/T^2$ as the
system enters the diffusive regime for $T>T_{\bf q}^{*}$. This
remarkable temperature dependence is a consequence of the fact that
the dissipation is proportional to the scattering time and thus
inversely proportional to the scattering rate.

In the presence of disorder, the temperature dependence also crosses
over  from constant to monotonically decreasing as $\propto 1/T^2$.
This crossover occurs at $T_{\bf q}^*$ when $q\ell \gg 1$ and at
$T^*$ when $q\ell \ll1$. When the synthetic electric field is purely
longitudinal, there can be an intermediate regime (for $T>T^*_{\bf
q}$ in the clean limit) in which the damping rate increases as $T^2$
before it crosses over into the $\propto 1/T^2$ behavior beyond a
temperature $(v_F/c)^{1/2}T^*_{\bf q}$.

\subsection{Damping by valley Coulomb drag in carbon nanotubes}

Valley Coulomb drag also affects the damping of phonon modes of
carbon nanotubes. For  the radial breathing mode of metallic carbon
nanotubes, the effect may be significant since phonon damping is
dominated by the decay of valley currents. Moreover, it is natural
to expect that Coulomb drag is particularly effective in a
one-dimensional setting. In contrast, we expect that valley Coulomb
drag is less significant for the longitudinal stretching mode where
damping is dominated by the effects of the scalar potential. In this
case, damping by decay of valley currents is only a subleading
contribution. Nevertheless, it is worthwhile to remark that valley
Coulomb drag provides the dominant damping mechanism based on the
electron-phonon interaction in the strictly clean limit.

A detailed theory of valley Coulomb drag in carbon nanotubes must
account for  possible electronic correlation effects associated with
the Luttinger liquid nature of the electron system. Such a theory is
beyond the scope of the present paper.

\section{Conclusions}
\label{conclusions}

We have considered the damping of low-energy phonons in carbon
nanotubes and  graphene originating from the electron-phonon
interaction. For most phonon modes, this damping is closely related
to a synthetic electric field associated with a strain-induced
vector potential in the Dirac equation for the electronic properties
of graphene. We find that it is very instructive to analyze phonon
damping in terms of these synthetic electric fields: (i) Within this
approach, phonon damping is a direct consequence of Joule heating.
(ii) This establishes a close relation between phonon damping and
the dynamic conductivity which we exploit to derive damping rates in
the presence of disorder and electron-electron interactions. (iii)
We find rich physics emerging from the fact that the synthetic
electric field has opposite signs in the two valleys. Most
prominently, we identify valley Coulomb drag as an important
dissipation channel which leads to unconventional temperature
dependence of the damping rate.

Throughout this paper, we have considered idealized samples in
the sense that we ignored finite size effects and electrodes.
Clearly, when suspended carbon nanotubes or graphene membranes are
coupled to electrodes, there will be (additional) electronic
dissipation taking place in the leads even if the nanotube is
otherwise perfectly ballistic. While the physics of this damping is
certainly highly non-universal, a rough estimate may be obtained
from our expressions for diffusive electronic dynamics by setting
the elastic mean free path equal to the length of the carbon
nanotube or the linear dimension of the graphene membrane. It is
also worthwhile to point out that the electron-phonon coupling is
not expected to change significantly in multilayer graphene samples.
Thus, our results should also be applicable in these systems.

Our results should be of direct relevance to the intense ongoing
experimental efforts to build and explore nanomechanical as well as
nanoelectromechanical devices based on graphene nanostructures. We
expect the electron-phonon interaction to be the dominant source of
phonon damping whenever the system exhibits a metallic conductivity.
In such systems, our results should be valuable by providing upper
bounds on the quality factor as well as by guiding optimization
strategies.

\begin{acknowledgments}
We would like to thank M. Polini for helpful discussions, and Y.\
Galperin for drawing our attention to the importance of intervalley
scattering. This work was supported in part by the Deutsche
Forschungsgemeinschaft through Sfb 658, SPP 1243, and DIP (FvO) by
MEC (Spain) through grants FIS2008-00124 and CONSOLIDER
CSD2007-00010, and by the Comunidad de Madrid, through CITECNOMIK
(FG). FvO and PG acknowledge the hopitality of the KITP at UCSB,
where research was supported in part by the National Science
Foundation under Grant No. PHY05-51164.
\end{acknowledgments}

\bigskip

\begin{appendix}

\section{Dynamic conductivity of graphene in the ballistic regime}

Here, we sketch the derivation of the longitudinal conductivity  in
the ballistic regime, including the effect of intervalley disorder
scattering. We start from Boltzmann equations with valley electric
field ${\bf E}$ for the two valleys,
\begin{eqnarray}
  &&(-i\omega + i {\bf v}_{\bf p}\cdot{\bf q} + \frac{e{\bf E}}{2}\cdot \nabla_{\bf p})n_{\bf p}^{(1)}\nonumber\\
     && \,\,\,\,\,\,\,\,\,\,\,\,\,
     = \frac{1}{\Omega\nu\tau_V}\sum_{{\bf p}^\prime} \delta(\epsilon_{\bf p}-\epsilon_{{\bf p}^\prime})
     (n_{{\bf p}^\prime}^{(2)}-n_{\bf p}^{(1))})\\
       &&(-i\omega + i {\bf v}_{\bf p}\cdot{\bf q} - \frac{e{\bf E}}{2}\cdot \nabla_{\bf p})n_{\bf p}^{(2)}\nonumber\\
     && \,\,\,\,\,\,\,\,\,\,\,\,\,
     = \frac{1}{\Omega\nu\tau_V}\sum_{{\bf p}^\prime} \delta(\epsilon_{\bf p}-\epsilon_{{\bf p}^\prime})
     (n_{{\bf p}^\prime}^{(1)}-n_{\bf p}^{(2))})
\end{eqnarray}
where $n_{\bf p}^{(j)}$ denotes the distribution function in valley
$j$.  Here we include the intervalley disorder scattering which is
required to obtain results which match the diffusive results. Taking
the difference between these two equations, we obtain an equation
for the odd distribution function $\Delta n_{\bf p}= n_{{\bf
p}}^{(1)}-n_{\bf p}^{(2))}$,
\begin{eqnarray}
  &&(-i\omega + i {\bf v}_{\bf p}\cdot{\bf q} + \frac{1}{\tau_V})\Delta n_{\bf p}
   + e{\bf E}\cdot\nabla_{\bf p}n_{\bf p}^{\rm (eq)}
   \nonumber\\
    && \,\,\,\,\,\,\,\,\,\,\,\,\,\,\,
     = -\frac{1}{\Omega\nu\tau_V}\sum_{{\bf p}^\prime} \delta(\epsilon_{\bf p}-\epsilon_{{\bf p}^\prime})
     \Delta n_{{\bf p}^\prime}.
\end{eqnarray}
Here, we asumed linear response and $n_{\bf p}^{\rm (eq)}$ denotes
the  Fermi-Dirac distribution. Introducing the Fermi surface
deformation $\delta\nu(\phi)$ by $\Delta n_{\bf p} =
\delta(\epsilon_{\bf p}-\mu)\delta\nu(\phi)$ and introducing the
angle $\phi=\angle({\bf q},{\bf p})=\angle({\bf E},{\bf p})$, we
find
\begin{equation}
  \delta \nu(\phi) = \frac{eEv_F\cos\phi - \frac{1}{\tau_V}\langle\delta\nu(\phi)\rangle}
    {-i\omega+iv_Fq\cos\phi +\frac{1}{\tau_V}}.
    \label{DeltaNu}
\end{equation}
In the absence of interlayer scattering, we find for the density
$\rho=e\nu\int(d\phi/2\pi)\delta\nu(\phi)$ the  conventional result
\begin{equation}
  \rho=\frac{e^2\nu}{iq}E .
\end{equation}
From the expression for the current, $j=e\nu v_F\int(d\phi/2\pi)
\cos\phi \delta\nu(\phi)$,  we obtain the (dissipative) longitudinal
conductivity
\begin{equation}
  \sigma_\parallel ({\bf q},\omega) = \frac{e^2\nu\omega^2}{v_Fq^3}.
\end{equation}

Eq.\ (\ref{DeltaNu}) shows that in the presence of intervalley
scattering,  there is an additional contribution (proportional to
$1/\tau_V$) to the current. In this term, we can evaluate $\langle
\delta\nu(\phi)\rangle$ using the result for $\delta\nu(\phi) \simeq
eE /iq$ in the absence of intervalley scattering. Noting that we may
also replace $\cos\phi$ in the first term by $\omega/v_Fq$, we find
that the second term dominates as long as $q\ll
[(q_0\ell_V)\ell]^{-1}$ while the first term dominates for $q\gg
[(q_0\ell_V)\ell]^{-1}$. Evaluating the longitudinal conductivity
then gives the results quoted in Eq.\ (\ref{LongCond}).

\end{appendix}

\bibliography{gauge_electric}

\begin{thebibliography}{30}
\expandafter\ifx\csname natexlab\endcsname\relax\def\natexlab#1{#1}\fi
\expandafter\ifx\csname bibnamefont\endcsname\relax
  \def\bibnamefont#1{#1}\fi
\expandafter\ifx\csname bibfnamefont\endcsname\relax
  \def\bibfnamefont#1{#1}\fi
\expandafter\ifx\csname citenamefont\endcsname\relax
  \def\citenamefont#1{#1}\fi
\expandafter\ifx\csname url\endcsname\relax
  \def\url#1{\texttt{#1}}\fi
\expandafter\ifx\csname urlprefix\endcsname\relax\def\urlprefix{URL }\fi
\providecommand{\bibinfo}[2]{#2}
\providecommand{\eprint}[2][]{\url{#2}}

\bibitem[{\citenamefont{Novoselov et~al.}(2004)\citenamefont{Novoselov, Geim,
  Morozov, Jiang, Zhang, Dubonos, Grigorieva, and Firsov}}]{Netal04}
\bibinfo{author}{\bibfnamefont{K.~S.} \bibnamefont{Novoselov}},
  \bibinfo{author}{\bibfnamefont{A.~K.} \bibnamefont{Geim}},
  \bibinfo{author}{\bibfnamefont{S.~V.} \bibnamefont{Morozov}},
  \bibinfo{author}{\bibfnamefont{D.}~\bibnamefont{Jiang}},
  \bibinfo{author}{\bibfnamefont{Y.}~\bibnamefont{Zhang}},
  \bibinfo{author}{\bibfnamefont{S.~V.} \bibnamefont{Dubonos}},
  \bibinfo{author}{\bibfnamefont{I.~V.} \bibnamefont{Grigorieva}},
  \bibnamefont{and} \bibinfo{author}{\bibfnamefont{A.~A.}
  \bibnamefont{Firsov}}, \bibinfo{journal}{Science}
  \textbf{\bibinfo{volume}{306}}, \bibinfo{pages}{666} (\bibinfo{year}{2004}).

\bibitem[{\citenamefont{Novoselov et~al.}(2005)\citenamefont{Novoselov, Jiang,
  Schedin, Booth, Khotkevich, Morozov, and Geim}}]{Netal05}
\bibinfo{author}{\bibfnamefont{K.~S.} \bibnamefont{Novoselov}},
  \bibinfo{author}{\bibfnamefont{D.}~\bibnamefont{Jiang}},
  \bibinfo{author}{\bibfnamefont{F.}~\bibnamefont{Schedin}},
  \bibinfo{author}{\bibfnamefont{T.~J.} \bibnamefont{Booth}},
  \bibinfo{author}{\bibfnamefont{V.~V.} \bibnamefont{Khotkevich}},
  \bibinfo{author}{\bibfnamefont{S.~V.} \bibnamefont{Morozov}},
  \bibnamefont{and} \bibinfo{author}{\bibfnamefont{A.~K.} \bibnamefont{Geim}},
  \bibinfo{journal}{Proc. Natl. Acad. Sci. U.S.A.}
  \textbf{\bibinfo{volume}{102}}, \bibinfo{pages}{10451}
  (\bibinfo{year}{2005}).

\bibitem[{\citenamefont{{Castro Neto} et~al.}(2009)\citenamefont{{Castro Neto},
  Guinea, Peres, Novoselov, and Geim}}]{NGPNG09}
\bibinfo{author}{\bibfnamefont{A.~H.} \bibnamefont{{Castro Neto}}},
  \bibinfo{author}{\bibfnamefont{F.}~\bibnamefont{Guinea}},
  \bibinfo{author}{\bibfnamefont{N.~M.~R.} \bibnamefont{Peres}},
  \bibinfo{author}{\bibfnamefont{K.~S.} \bibnamefont{Novoselov}},
  \bibnamefont{and} \bibinfo{author}{\bibfnamefont{A.~K.} \bibnamefont{Geim}},
  \bibinfo{journal}{Rev. Mod. Phys.} \textbf{\bibinfo{volume}{81}},
  \bibinfo{pages}{109} (\bibinfo{year}{2009}).

\bibitem[{\citenamefont{Gonz\'alez et~al.}(1992)\citenamefont{Gonz\'alez,
  Guinea, and Vozmediano}}]{GGV92}
\bibinfo{author}{\bibfnamefont{J.}~\bibnamefont{Gonz\'alez}},
  \bibinfo{author}{\bibfnamefont{F.}~\bibnamefont{Guinea}}, \bibnamefont{and}
  \bibinfo{author}{\bibfnamefont{M.~A.~H.} \bibnamefont{Vozmediano}},
  \bibinfo{journal}{Phys. Rev. Lett.} \textbf{\bibinfo{volume}{69}},
  \bibinfo{pages}{172} (\bibinfo{year}{1992}).

\bibitem[{\citenamefont{Suzuura and Ando}(2002)}]{SA02b}
\bibinfo{author}{\bibfnamefont{H.}~\bibnamefont{Suzuura}} \bibnamefont{and}
  \bibinfo{author}{\bibfnamefont{T.}~\bibnamefont{Ando}},
  \bibinfo{journal}{Phys. Rev. B} \textbf{\bibinfo{volume}{65}},
  \bibinfo{pages}{235412} (\bibinfo{year}{2002}).

\bibitem[{\citenamefont{Morozov et~al.}(2006)\citenamefont{Morozov, Novoselov,
  Katsnelson, Schedin, Ponomarenko, Jiang, and Geim}}]{Metal06}
\bibinfo{author}{\bibfnamefont{S.~V.} \bibnamefont{Morozov}},
  \bibinfo{author}{\bibfnamefont{K.~S.} \bibnamefont{Novoselov}},
  \bibinfo{author}{\bibfnamefont{M.~I.} \bibnamefont{Katsnelson}},
  \bibinfo{author}{\bibfnamefont{F.}~\bibnamefont{Schedin}},
  \bibinfo{author}{\bibfnamefont{L.~A.} \bibnamefont{Ponomarenko}},
  \bibinfo{author}{\bibfnamefont{D.}~\bibnamefont{Jiang}}, \bibnamefont{and}
  \bibinfo{author}{\bibfnamefont{A.~K.} \bibnamefont{Geim}},
  \bibinfo{journal}{Phys. Rev. Lett.} \textbf{\bibinfo{volume}{97}},
  \bibinfo{pages}{016801} (\bibinfo{year}{2006}).

\bibitem[{\citenamefont{{Ma\~nes}}(2007)}]{M07}
\bibinfo{author}{\bibfnamefont{J.~L.} \bibnamefont{{Ma\~nes}}},
  \bibinfo{journal}{Phys. Rev. B} \textbf{\bibinfo{volume}{76}},
  \bibinfo{pages}{045430} (\bibinfo{year}{2007}).

\bibitem[{\citenamefont{Morpurgo and Guinea}(2006)}]{MG06}
\bibinfo{author}{\bibfnamefont{A.}~\bibnamefont{Morpurgo}} \bibnamefont{and}
  \bibinfo{author}{\bibfnamefont{F.}~\bibnamefont{Guinea}},
  \bibinfo{journal}{Phys. Rev. Lett.} \textbf{\bibinfo{volume}{97}},
  \bibinfo{pages}{196804} (\bibinfo{year}{2006}).

\bibitem[{\citenamefont{Guinea et~al.}(2008{\natexlab{a}})\citenamefont{Guinea,
  Katsnelson, and Vozmediano}}]{GKV08}
\bibinfo{author}{\bibfnamefont{F.}~\bibnamefont{Guinea}},
  \bibinfo{author}{\bibfnamefont{M.~I.} \bibnamefont{Katsnelson}},
  \bibnamefont{and} \bibinfo{author}{\bibfnamefont{M.~A.~H.}
  \bibnamefont{Vozmediano}}, \bibinfo{journal}{Phys. Rev. B}
  \textbf{\bibinfo{volume}{77}}, \bibinfo{pages}{075422}
  (\bibinfo{year}{2008}{\natexlab{a}}).

\bibitem[{\citenamefont{Mariani and von Oppen}(2008)}]{MO08}
\bibinfo{author}{\bibfnamefont{E.}~\bibnamefont{Mariani}} \bibnamefont{and}
  \bibinfo{author}{\bibfnamefont{F.}~\bibnamefont{von Oppen}},
  \bibinfo{journal}{Phys. Rev. Lett.} \textbf{\bibinfo{volume}{100}},
  \bibinfo{pages}{076801} (\bibinfo{year}{2008}).

\bibitem[{\citenamefont{Guinea et~al.}(2008{\natexlab{b}})\citenamefont{Guinea,
  Horovitz, and Doussal}}]{GHL08}
\bibinfo{author}{\bibfnamefont{F.}~\bibnamefont{Guinea}},
  \bibinfo{author}{\bibfnamefont{B.}~\bibnamefont{Horovitz}}, \bibnamefont{and}
  \bibinfo{author}{\bibfnamefont{P.~L.} \bibnamefont{Doussal}},
  \bibinfo{journal}{Phys. Rev. B} \textbf{\bibinfo{volume}{77}},
  \bibinfo{pages}{205421} (\bibinfo{year}{2008}{\natexlab{b}}).

\bibitem[{\citenamefont{Sasaki et~al.}(2008)\citenamefont{Sasaki, Saito,
  Dresselhaus, Dresselhaus, Farhat, and Kong}}]{Setal08b}
\bibinfo{author}{\bibfnamefont{K.}~\bibnamefont{Sasaki}},
  \bibinfo{author}{\bibfnamefont{R.}~\bibnamefont{Saito}},
  \bibinfo{author}{\bibfnamefont{G.}~\bibnamefont{Dresselhaus}},
  \bibinfo{author}{\bibfnamefont{M.~S.} \bibnamefont{Dresselhaus}},
  \bibinfo{author}{\bibfnamefont{H.}~\bibnamefont{Farhat}}, \bibnamefont{and}
  \bibinfo{author}{\bibfnamefont{J.}~\bibnamefont{Kong}},
  \bibinfo{journal}{Phys. Rev. B} \textbf{\bibinfo{volume}{78}},
  \bibinfo{pages}{235405} (\bibinfo{year}{2008}).

\bibitem[{\citenamefont{Martino et~al.}(2009)\citenamefont{Martino, Egger, and
  Gogolin}}]{MEG09}
\bibinfo{author}{\bibfnamefont{A.~D.} \bibnamefont{Martino}},
  \bibinfo{author}{\bibfnamefont{R.}~\bibnamefont{Egger}}, \bibnamefont{and}
  \bibinfo{author}{\bibfnamefont{A.~O.} \bibnamefont{Gogolin}}
  (\bibinfo{year}{2009}), \eprint{arXiv:0903.1771}.

\bibitem[{\citenamefont{{D'Amico} and Vignale}(2001)}]{Vig01}
\bibinfo{author}{\bibfnamefont{I.}~\bibnamefont{{D'Amico}}} \bibnamefont{and}
  \bibinfo{author}{\bibfnamefont{G.}~\bibnamefont{Vignale}},
  \bibinfo{journal}{Europhys. Lett.} \textbf{\bibinfo{volume}{55}},
  \bibinfo{pages}{566} (\bibinfo{year}{2001}).

\bibitem[{\citenamefont{Flensberg et~al.}(2001)\citenamefont{Flensberg, Jensen,
  and Mortensen}}]{Fle01}
\bibinfo{author}{\bibfnamefont{K.}~\bibnamefont{Flensberg}},
  \bibinfo{author}{\bibfnamefont{T.~S.} \bibnamefont{Jensen}},
  \bibnamefont{and} \bibinfo{author}{\bibfnamefont{N.~A.}
  \bibnamefont{Mortensen}}, \bibinfo{journal}{Phys. Rev. B}
  \textbf{\bibinfo{volume}{64}}, \bibinfo{pages}{245308}
  (\bibinfo{year}{2001}).

\bibitem[{\citenamefont{Weber et~al.}(2005)\citenamefont{Weber, Gedik, Moore,
  Orenstein, Stephens, and Awschalom}}]{Web05}
\bibinfo{author}{\bibfnamefont{C.~P.} \bibnamefont{Weber}},
  \bibinfo{author}{\bibfnamefont{N.}~\bibnamefont{Gedik}},
  \bibinfo{author}{\bibfnamefont{J.~E.} \bibnamefont{Moore}},
  \bibinfo{author}{\bibfnamefont{J.}~\bibnamefont{Orenstein}},
  \bibinfo{author}{\bibfnamefont{J.}~\bibnamefont{Stephens}}, \bibnamefont{and}
  \bibinfo{author}{\bibfnamefont{D.~D.} \bibnamefont{Awschalom}},
  \bibinfo{journal}{Nature} \textbf{\bibinfo{volume}{437}},
  \bibinfo{pages}{1330} (\bibinfo{year}{2005}).

\bibitem[{\citenamefont{LeRoy et~al.}(2004)\citenamefont{LeRoy, Lemay, Kong,
  and Dekker}}]{LLKD04}
\bibinfo{author}{\bibfnamefont{B.~J.} \bibnamefont{LeRoy}},
  \bibinfo{author}{\bibfnamefont{S.~G.} \bibnamefont{Lemay}},
  \bibinfo{author}{\bibfnamefont{J.}~\bibnamefont{Kong}}, \bibnamefont{and}
  \bibinfo{author}{\bibfnamefont{C.}~\bibnamefont{Dekker}},
  \bibinfo{journal}{Nature} \textbf{\bibinfo{volume}{432}},
  \bibinfo{pages}{371} (\bibinfo{year}{2004}).

\bibitem[{\citenamefont{Rao et~al.}(2007)\citenamefont{Rao, Menendez, Poweleit,
  and Rao}}]{RMPR06}
\bibinfo{author}{\bibfnamefont{R.}~\bibnamefont{Rao}},
  \bibinfo{author}{\bibfnamefont{J.}~\bibnamefont{Menendez}},
  \bibinfo{author}{\bibfnamefont{C.~D.} \bibnamefont{Poweleit}},
  \bibnamefont{and} \bibinfo{author}{\bibfnamefont{A.~M.} \bibnamefont{Rao}},
  \bibinfo{journal}{Phys. Rev. Lett.} \textbf{\bibinfo{volume}{99}},
  \bibinfo{pages}{047403} (\bibinfo{year}{2007}).

\bibitem[{\citenamefont{Sapmaz et~al.}(2006)\citenamefont{Sapmaz,
  Jarillo-Herrero, Blanter, Dekker, and van~der Zant}}]{Setal06}
\bibinfo{author}{\bibfnamefont{S.}~\bibnamefont{Sapmaz}},
  \bibinfo{author}{\bibfnamefont{P.}~\bibnamefont{Jarillo-Herrero}},
  \bibinfo{author}{\bibfnamefont{Y.~M.} \bibnamefont{Blanter}},
  \bibinfo{author}{\bibfnamefont{C.}~\bibnamefont{Dekker}}, \bibnamefont{and}
  \bibinfo{author}{\bibfnamefont{H.~S.~J.} \bibnamefont{van~der Zant}},
  \bibinfo{journal}{Phys. Rev. Lett.} \textbf{\bibinfo{volume}{96}},
  \bibinfo{pages}{026801} (\bibinfo{year}{2006}).

\bibitem[{\citenamefont{Witkamp et~al.}(2006)\citenamefont{Witkamp, Poot, and
  van~der Zant}}]{WPV06}
\bibinfo{author}{\bibfnamefont{B.}~\bibnamefont{Witkamp}},
  \bibinfo{author}{\bibfnamefont{M.}~\bibnamefont{Poot}}, \bibnamefont{and}
  \bibinfo{author}{\bibfnamefont{H.~S.~J.} \bibnamefont{van~der Zant}},
  \bibinfo{journal}{Nano Lett.} \textbf{\bibinfo{volume}{6}},
  \bibinfo{pages}{2904} (\bibinfo{year}{2006}).

\bibitem[{\citenamefont{H{\"u}ttel et~al.}(2008)\citenamefont{H{\"u}ttel, Poot,
  Witkamp, and {S. J. van der Zant}}}]{HPWZ08}
\bibinfo{author}{\bibfnamefont{A.~K.} \bibnamefont{H{\"u}ttel}},
  \bibinfo{author}{\bibfnamefont{M.}~\bibnamefont{Poot}},
  \bibinfo{author}{\bibfnamefont{B.}~\bibnamefont{Witkamp}}, \bibnamefont{and}
  \bibinfo{author}{\bibfnamefont{H.}~\bibnamefont{{S. J. van der Zant}}},
  \bibinfo{journal}{New Journ. Phys.} \textbf{\bibinfo{volume}{10}},
  \bibinfo{pages}{095003} (\bibinfo{year}{2008}).

\bibitem[{\citenamefont{Leturcq et~al.}(2009)\citenamefont{Leturcq, Stampfer,
  Inderbitzin, Durrer, Hierold, Mariani, Schultz, von Oppen, and
  Ensslin}}]{Letal09}
\bibinfo{author}{\bibfnamefont{R.}~\bibnamefont{Leturcq}},
  \bibinfo{author}{\bibfnamefont{C.}~\bibnamefont{Stampfer}},
  \bibinfo{author}{\bibfnamefont{K.}~\bibnamefont{Inderbitzin}},
  \bibinfo{author}{\bibfnamefont{L.}~\bibnamefont{Durrer}},
  \bibinfo{author}{\bibfnamefont{C.}~\bibnamefont{Hierold}},
  \bibinfo{author}{\bibfnamefont{E.}~\bibnamefont{Mariani}},
  \bibinfo{author}{\bibfnamefont{M.~G.} \bibnamefont{Schultz}},
  \bibinfo{author}{\bibfnamefont{F.}~\bibnamefont{von Oppen}},
  \bibnamefont{and} \bibinfo{author}{\bibfnamefont{K.}~\bibnamefont{Ensslin}},
  \bibinfo{journal}{Nature Physics}  (\bibinfo{year}{2009}), \bibinfo{note}{(in
  press)}.

\bibitem[{\citenamefont{Garcia-Sanchez
  et~al.}(2007)\citenamefont{Garcia-Sanchez, Paulo, Esplandiu, Perez-Murano,
  Forr{\'o}, Aguasca, and Bachtold}}]{Getal07b}
\bibinfo{author}{\bibfnamefont{D.}~\bibnamefont{Garcia-Sanchez}},
  \bibinfo{author}{\bibfnamefont{A.~S.} \bibnamefont{Paulo}},
  \bibinfo{author}{\bibfnamefont{M.~J.} \bibnamefont{Esplandiu}},
  \bibinfo{author}{\bibfnamefont{F.}~\bibnamefont{Perez-Murano}},
  \bibinfo{author}{\bibfnamefont{L.}~\bibnamefont{Forr{\'o}}},
  \bibinfo{author}{\bibfnamefont{A.}~\bibnamefont{Aguasca}}, \bibnamefont{and}
  \bibinfo{author}{\bibfnamefont{A.}~\bibnamefont{Bachtold}},
  \bibinfo{journal}{Phys. Rev. Lett.} \textbf{\bibinfo{volume}{99}},
  \bibinfo{pages}{085501} (\bibinfo{year}{2007}).

\bibitem[{\citenamefont{Bunch et~al.}(2007)\citenamefont{Bunch, {van der
  Zande}, Verbridge, Frank, Tanenbaum, Parpia, Craighead, and
  McEuen}}]{Betal07}
\bibinfo{author}{\bibfnamefont{J.~S.} \bibnamefont{Bunch}},
  \bibinfo{author}{\bibfnamefont{A.~M.} \bibnamefont{{van der Zande}}},
  \bibinfo{author}{\bibfnamefont{S.~S.} \bibnamefont{Verbridge}},
  \bibinfo{author}{\bibfnamefont{I.~W.} \bibnamefont{Frank}},
  \bibinfo{author}{\bibfnamefont{D.~M.} \bibnamefont{Tanenbaum}},
  \bibinfo{author}{\bibfnamefont{J.~M.} \bibnamefont{Parpia}},
  \bibinfo{author}{\bibfnamefont{H.~G.} \bibnamefont{Craighead}},
  \bibnamefont{and} \bibinfo{author}{\bibfnamefont{P.~L.}
  \bibnamefont{McEuen}}, \bibinfo{journal}{Science}
  \textbf{\bibinfo{volume}{315}}, \bibinfo{pages}{5811} (\bibinfo{year}{2007}).

\bibitem[{\citenamefont{Garcia-Sanchez
  et~al.}(2008)\citenamefont{Garcia-Sanchez, {van der Zande}, Paulo, Lassagne,
  McEuen, and Bachtold}}]{Getal08}
\bibinfo{author}{\bibfnamefont{D.}~\bibnamefont{Garcia-Sanchez}},
  \bibinfo{author}{\bibfnamefont{A.~M.} \bibnamefont{{van der Zande}}},
  \bibinfo{author}{\bibfnamefont{A.~S.} \bibnamefont{Paulo}},
  \bibinfo{author}{\bibfnamefont{B.}~\bibnamefont{Lassagne}},
  \bibinfo{author}{\bibfnamefont{P.~L.} \bibnamefont{McEuen}},
  \bibnamefont{and} \bibinfo{author}{\bibfnamefont{A.}~\bibnamefont{Bachtold}},
  \bibinfo{journal}{Nano Lett.} \textbf{\bibinfo{volume}{8}},
  \bibinfo{pages}{1399} (\bibinfo{year}{2008}).

\bibitem[{\citenamefont{Koch and von Oppen}(2005)}]{KO05}
\bibinfo{author}{\bibfnamefont{J.}~\bibnamefont{Koch}} \bibnamefont{and}
  \bibinfo{author}{\bibfnamefont{F.}~\bibnamefont{von Oppen}},
  \bibinfo{journal}{Phys. Rev. Lett.} \textbf{\bibinfo{volume}{94}},
  \bibinfo{pages}{206804} (\bibinfo{year}{2005}).

\bibitem[{\citenamefont{Koch et~al.}(2006)\citenamefont{Koch, von Oppen, and
  Andreev}}]{KOA06}
\bibinfo{author}{\bibfnamefont{J.}~\bibnamefont{Koch}},
  \bibinfo{author}{\bibfnamefont{F.}~\bibnamefont{von Oppen}},
  \bibnamefont{and} \bibinfo{author}{\bibfnamefont{A.}~\bibnamefont{Andreev}},
  \bibinfo{journal}{Phys. Rev. B} \textbf{\bibinfo{volume}{74}},
  \bibinfo{pages}{205438} (\bibinfo{year}{2006}).

\bibitem[{\citenamefont{Pomerantz et~al.}(1962)\citenamefont{Pomerantz, Keyes,
  and Seiden}}]{Pomerantz62}
\bibinfo{author}{\bibfnamefont{M.}~\bibnamefont{Pomerantz}},
  \bibinfo{author}{\bibfnamefont{R.~W.} \bibnamefont{Keyes}}, \bibnamefont{and}
  \bibinfo{author}{\bibfnamefont{P.}~\bibnamefont{Seiden}},
  \bibinfo{journal}{Phys. Rev. Lett.} \textbf{\bibinfo{volume}{9}},
  \bibinfo{pages}{312} (\bibinfo{year}{1962}).

\bibitem[{\citenamefont{Halperin et~al.}(1993)\citenamefont{Halperin, Lee, and
  Read}}]{Halperin}
\bibinfo{author}{\bibfnamefont{B.}~\bibnamefont{Halperin}},
  \bibinfo{author}{\bibfnamefont{P.}~\bibnamefont{Lee}}, \bibnamefont{and}
  \bibinfo{author}{\bibfnamefont{N.}~\bibnamefont{Read}},
  \bibinfo{journal}{Phys. Rev. B} \textbf{\bibinfo{volume}{47}},
  \bibinfo{pages}{7312} (\bibinfo{year}{1993}).

\bibitem[{\citenamefont{Tse et~al.}(2007)\citenamefont{Tse, Hu, and
  Sarma}}]{THS07}
\bibinfo{author}{\bibfnamefont{W.-K.} \bibnamefont{Tse}},
  \bibinfo{author}{\bibfnamefont{Y.-K.} \bibnamefont{Hu}}, \bibnamefont{and}
  \bibinfo{author}{\bibfnamefont{S.~D.} \bibnamefont{Sarma}},
  \bibinfo{journal}{Phys. Rev. B} \textbf{\bibinfo{volume}{76}},
  \bibinfo{pages}{081401} (\bibinfo{year}{2007}).

\end{thebibliography}
\end{document}